\documentclass[9pt,twocolumn,twoside]{osajnl}

\journal{ol} 

\setboolean{shortarticle}{true}

\title{Pulse overlap ambiguities in multiple quantum coherence spectroscopy}

\author[1]{Ulrich Bangert}
\author[*,1]{Lukas Bruder}%
\author[1]{Frank Stienkemeier}

\affil{Institute of Physics, University of Freiburg, Hermann-Herder-Str. 3, 79104 Freiburg, Germany.}

\affil[*]{Corresponding author: lukas.bruder@physik.uni-freiburg.de}

\begin{abstract}
  Coherent two-dimensional electronic spectroscopy probes ultrafast dynamics using femtosecond pulses.
  In case the timescale of the studied dynamics become comparable to the pulse duration, pulse overlap effects may compromise the experimental data.
  Here, we perform one-dimensional coherence scans and study pulse overlap effects in clean two-level systems. 
  We find parasitic multiple-quantum coherences as a consequence of the arbitrary time ordering during the temporal pulse overlap.
  Surprisingly, the coherence lifetimes exceed the pulse coherence time by a factor of 1.85. 
  These findings have important implications for the interpretation of higher-order coherent two-dimensional and related spectroscopy experiments. 
\end{abstract}

\setboolean{displaycopyright}{true}

\begin{document}

\maketitle

\begin{figure}[htbp]
\centering
\fbox{\includegraphics{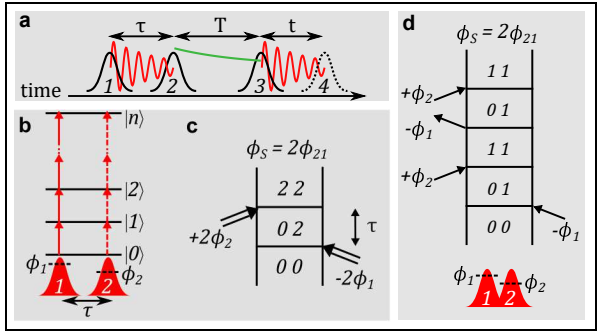}}
\caption{(a) Pulse sequence in 2DES showing the decomposition into 1D coherence scans (red) and the free evolution of the system (green). 
(b) Schematic of a multi quantum coherence two-pulse excitation in an n-level system, details in text. 
(c) Feynman diagram of standard 2QC induced by two temporally separated pulses. Arrows indicate interactions with pulse $i$, imprinting its phase $\phi_i$. 
(d) Exemplary Feynman diagram of artificial 2QC pathways occuring in the pulse-overlap regime. 
The time-ordering of the light-matter interactions are no longer well-defined and  only singly excited states are required to generate a 2QC signal signature. 
$\phi_S$ denotes the phase signature of the respective signals.}
\label{fig1}
\end{figure}

Two-dimensional  electronic spectroscopy (2DES) achieves a high spectro-temporal resolution by combining femtosecond laser technology with interferometric measurement schemes\,\cite{jonas_two-dimensional_2003, fuller_experimental_2015}. 
To this end, the sample is excited with a sequence of 3 to 4 femtosecond pulses and the nonlinear response is recorded as a function of the inter-pulse delays (Fig.\,\ref{fig1}a). 
The experimental routine can be decomposed into (i) one-dimensional (1D) coherence scans (delays $\tau, t$) which track the time evolution of electronic coherences between different electronic states, and (ii) the free evolution of the system in between the coherence scans (delay $T$), probing the dynamics of the system. 
A Fourier transform of the signal provides multidimensional spectra which directly discloses the frequency correlations in the nonlinear system response\,\cite{note1}.
If the electronic coherence probed during the coherence scans, is induced by a one-photon transition, the corresponding signal is termed single-quantum coherence (1QC). For multi-photon excitations the signal is termed multiple or $n$-quantum coherence (MQC or nQC, with $n \in \mathbb{N}$), respectively (see Fig.\,\ref{fig1}).
MQC signals can be used to readily probe higher lying states and as a sensitive probe of intra and inter-particle couplings in various systems. 
As such, 2DES and related experiments involving MQC signals became increasingly popular in recent years\, \cite{bruder_efficient_2015,kim_two-dimensional_2009, christensson_electronic_2010,karaiskaj_two-quantum_2010,mueller_observing_2021,bruder_delocalized_2019,dai_two-dimensional_2012}.
While these experiments focused on the study of double quantum coherences (2QCs), also higher-order nQC ($n>2$) signals where investigated\,\cite{turner_coherent_2010, dostal_direct_2018,heshmatpour_annihilation_2020, kriete_interplay_2019, yu_long_2019}.

The inherent timescales of MQC experiments are often fast, as in highly excited systems the high density of states and ultrafast internal conversion channels increase the dephasing and decay rates. Consequently, the signal lifetimes can approach the finite pulse duration of the excitation pulses. 
In this case, the interpretation of 2DES results can become biased when evaluated in and close to the temporal pulse overlap. In particular, the non-existing time ordering during the pulse overlap leads to mixing of different excitation pathways that are else separated by phase matching or phase cycling conditions\cite{maly_coherently_2020,palecek_potential_2019,fumero_resolution_2015}. 
Previous studies\,\cite{maly_coherently_2020,palecek_potential_2019} focused on the possible ambiguities caused by overlapping pulses 2 and 3, thus, occurring for short evolution times $T$.
In this letter, we show that in nQC signals parasitic pulse overlap effects contribute also for the temporal overlap in the coherence scan.
To this end we study a clean two-level system in the gas phase with no coupling to higher-lying states or neighboring particles. Hence, genuine MQC signals are not present in these systems which enables us to unambiguously identify parasitic pulse overlap features leading to artificial MQC signals.
Fourth-order runge-kutta simulations further confirm our observations, thus, excluding experimental artifacts as an origin. 
To put the duration of the artificial MQC signal into relation, we compare it to a purely optical coherence and the intensity autocorrelation (AC) of the excitation pulses.
The former estimates a limiting case for short-lived coherences and the latter directly quantifies the pulse overlap.

Since we focus our study on pulse-overlap artifacts occurring during the coherence scans, we reduce the 2DES experiment to a single 1D coherence measurement. 
Phase modulation\,\cite{tekavec_wave_2006} is applied to extract the nonlinear signals with an MQC-detection scheme described in Ref.\,\cite{bruder_efficient_2015}. It is straightforward to extend this approach to multidimensional schemes\,\cite{dai_two-dimensional_2012, yu_long_2019}. 
The experimental principle is shown in Fig.\,\ref{fig1}b, c and briefly discussed in the following. The first pulse generates a MQC and the second pulse maps it into a population state, both upon multiple interactions of the respective pulse with the target system. The evolution of the MQC is measured by scanning the inter-pulse delay $\tau$ and detecting the change in population via fluorescence detection. 
The phase $\phi_i$ ($i$=1,2) of each pulse is modulated on a shot-to-shot basis. 
Each light-matter interaction imprints this phase onto the system leading to a quasi-continuous real-time modulation of the fluorescence signal according to the imprinted phase signature $\phi_S=\sum_{m,n} a_m \phi_1 + b_n \phi_2$ ($a_m, b_n = \pm 1$) and thus its modulation frequency depends on the interaction sequence.
To map the MQC into a population state, an identical number of interactions with each pulse is required ($n=m$) and further the signs of the interaction have to be opposite ($a_m = - b_n$). Note, we only consider the lowest order here, as it is the dominant contribution.
Consequently, the nQC signals occur at the n'th harmonic modulation frequency of the 1QC signal ($\phi_{21} = \phi_2-\phi_1$) and are conveniently detected by e.g. harmonic lock-in detection\,\cite{bruder_efficient_2015}. In this study, we are confined to 1QC and 2QCs due to low signal yields.
We note, that the phase modulation approach is closely related to phase cycling MQC experiments\, \cite{mueller_rapid_2019} and our results are expected to be equally valid for these type of experiments. 

As samples, we use a highly dilute lithium (Li) vapor (particle density $\sim 10^7$ cm$^{-3}$) and freebase phthalocyanine molecules isolated in helium nanodroplets (H$_2$Pc-He$_N$) (details in Ref.\,\cite{bangert_high-resolution_2022}). Both systems feature isolated one-photon transitions (Li: $2s \rightarrow 2p$, 670.8\,nm, H$_2$Pc: $\mathrm{S}_0 \rightarrow \mathrm{S}_1$, 662.3\,nm) without any coupling to higher lying states in the range of our laser spectrum (shown in Fig.\,\ref{fig3}). Any collective excitations are further excluded by the preparation of our sample. In the Li vapor, collective excitations\,\cite{bruder_delocalized_2019} are suppressed below the noise level by choosing a low vapor density. The H$_2$Pc molecules are isolated in superfluid heliumdroplets, which serve as nano-containers isolating each H$_2$Pc molecule and thus suppressing inter-molecular interactions in the ensemble comparable to the model of a “frozen” highly dilute molecular beam\,\cite{toennies_superfluid_2004}. Hence, both systems are ideal two level systems for which one would only expect 1QC signals.
We use the following laser parameters: center wavelength: 668.5\,nm, spectral FWHM:  25\,nm, pulse duration: 47\,fs, pulse energies: 34\,nJ, focus diameter: 200\,$\mu$m.
The MQC signals and the simulated data are sampled at steps of 10\,fs, the optical 2QC signal at steps of 5\,fs.

 \begin{figure}[htb]
\centering
\fbox{\includegraphics{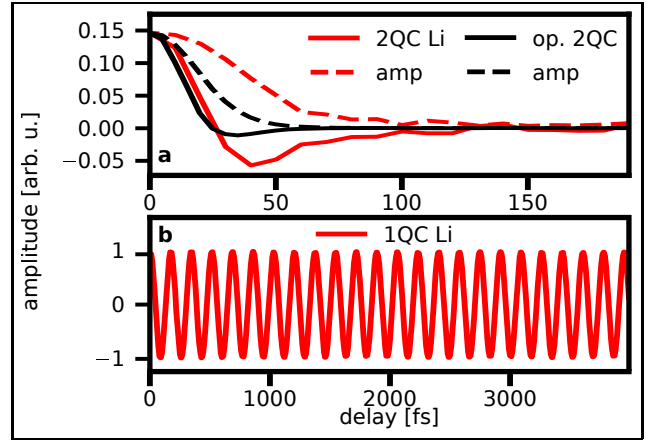}}
\caption{Time-domain 2QC in (a) and 1QC signal in (b) detected in a dilute Li atom vapor (red). 
Optical 2QC of the excitation pulses (black). 
Solid lines show the real part of the complex-valued signals, exhibiting clear coherent oscillations. 
Dashed lines show the absolute values, reflecting the amplitude decay of the signals. 
For better comparison, the 2QC signals are normalized to the 1QC signal, the optical 2QC amplitude is scaled to the Li 2QC signal amplitude. The oscillation frequencies of all signals are downshifted due to rotating frame detection\,\cite{tekavec_wave_2006}.}
\label{fig2}
\end{figure}

Despite the clean two level systems, we can clearly observe MQC signals in the experiment.
Fig.\,\ref{fig2} shows the 1QC and 2QC recorded in Li. The 1QC has a very long lifetime.
It decays within a few nanoseconds due to Doppler broadening which is far beyond our scan range.
In stark contrast, the 2QC rapidly decays on a time scale in the order of the pulse duration.
Note that MQC signals from collective excitations  are commonly observed in alkaline vapors even at low densities\,\cite{yu_long_2019}.
But their longevity clearly distinguishes them from the features observed here.

As an estimated lower limit for short-lived coherences in the 2QC channel, we generate an optical 2QC  using second harmonic generation (SHG) in a beta barium borate crystal (crystal thickness: 10\,$\mu$m).
The SHG light of the collinear pulse pair is recorded with a photo diode and evaluated according to the phase modulation detection scheme.
This measurement  corresponds to a second order interferometric AC which can be decomposed into three contributions: the intensity AC, a 1f and a 2f frequency component \cite{trager2012springer}.
The phase modulation scheme explicitly picks out the 2f-contribution which corresponds to the first-order interferometric AC of the SHG pulses.
In the remainder, this signal is termed optical 2QC, as its coherence behaviour is given by the optical pulses.
The comparison between both signals reveals that the lifetime of the 2QC in Li is a factor of 1.85 longer than the optical 2QC.
Qualitatively the same behavior is found for the H$_2$Pc-He$_N$ system (not shown).
Compared to the intensity AC of the pulses the Li 2QC lives longer by a factor of 1.15.

 \begin{figure}[htb]
\centering
\fbox{\includegraphics{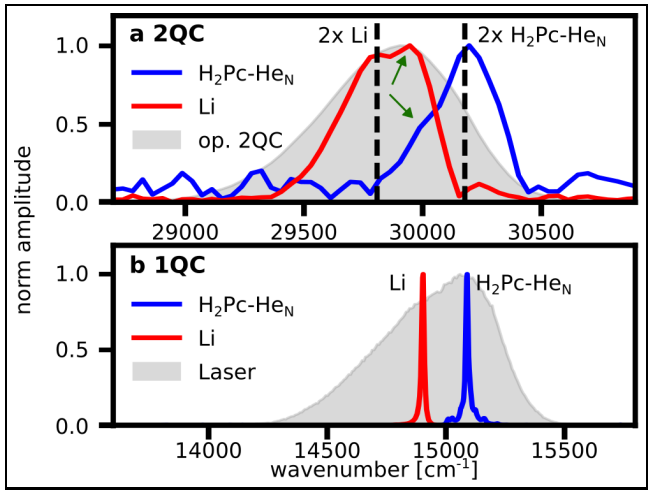}}
\caption{Fourier transform of the 1QCs (in (b)) and 2QCs (in (a)) in Li and H$_2$Pc-He$_N$. Dashed black lines indicate the position of twice the single-photon resonance frequency of the systems. The fundamental and SHG pulse spectrum for (b) and (a), respectively, are shown in grey. The SHG spectrum is obtained from Fourier transform of the optical 2QC. The Li and H$_2$Pc-He$_N$ data were recorded in separate experiments. 
All 1QC and 2QC signal amplitudes are normalized to unity. 
}
\label{fig3}
\end{figure}

Fig.\,\ref{fig3} shows the Fourier transform of the signals with respect to $\tau$. 
The 1QC spectra of Li and H$_2$Pc-He$_N$ feature a single dominant narrow absorption line as expected for a pure two level system. The linewidth is here determined by the experimental resolution (12\,cm$^{-1}$) which is much broader than the actual linewidth of the 1QC signals (<0.3\,cm$^{-1}$). Note that the Li spectrum actually consist of the D1- and D2-line, which are not spectrally resolved. The additional weak spectral lines in the H$_2$Pc-He$_N$ originate from sample impurities and do not interact with the main line, as shown recently\,\cite{bangert_high-resolution_2022}, thus they will not induce 2QC signals in the system.
The Fourier transform of the 2QC signals show a broad double peak structure for both samples which are fully resolved within the experimental resolution. 
One of the sub-peaks correlates to twice the one-photon absorption frequencies (indicated by the dashed lines) of the respective sample and the second sub-peak (marked by the arrows) is shifted towards twice the center frequency of the optical pulses as indicated by the Fourier spectrum of the optical 2QC. 

Assuming a well-defined time ordering of the pulses, short-lived 2QC signals should not appear for the clean and well-isolated two level systems studied here. 
Moreover, the spectral width and the structured lineshapes of the 2QC signals are in clear contrast to the Fourier transform of the optical 2QC and thus cannot be explained solely by optical frequency mixing processes. 
Instead, the 2QC signals can be qualitatively explained by Raman-like processes which may occur during the temporal overlap of the pulses. 
Fig.\,\ref{fig1} summarizes the situation. 
Considering strictly time-ordered interactions, a 2QC signal can only arise from a coherence between ground and second excited state excited by sequential double-interactions with either of the two pulses (Fig.\,\ref{fig1}c). 
The phase signature of the 2QC signal is $\phi_\mathrm{S}=2\phi_{21}$. 
During the pulse overlap, however, the pulses can alternately interact with the system as shown in Fig.\,\ref{fig1}d.
These new pathways exhibit the same phase signature as a 2QC signal, but only involve the excitation into the first excited state.
Hence, they also exist in a pure two level system. 
These pathways, thus, provide a possible explanation for the occurrence of 2QC signals in our samples.
A quantitative discussion of the signal contributions based on double-sided Feynman diagrams becomes less intuitive, as the timing and time-ordering of the light-matter interactions are not well-defined during pulse overlap.
Here, we employ a non-perturbative numerical calculation of the signal in order to account for all effects on a quantitative level.

\begin{figure}[htbp]
\centering
\fbox{\includegraphics{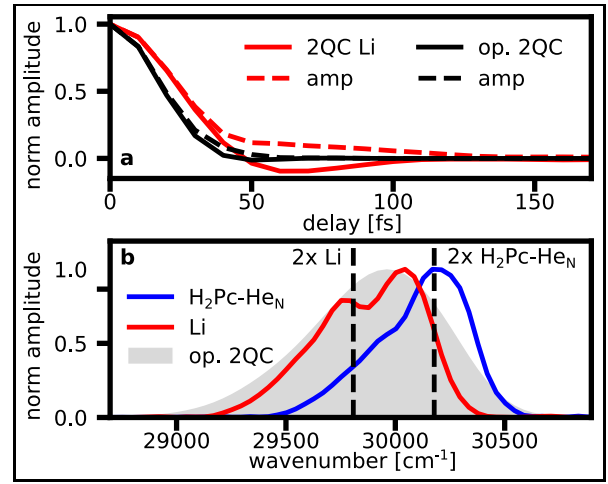}}
\caption{Simulated 2QC signals for a pulse peak intensity of 0.1\,GW/cm$^2$. (a) Temporal evolution of 2QC in Li and of optical 2QC. (b) Fourier transformation of 2QC in Li and in H$_2$Pc-He$_N$ and of optical 2QC. Representations are analog to Fig.\,\ref{fig2} \& Fig\,\ref{fig3}. }
\label{fig4}
\end{figure}

The non-perturbative calculations are based on a fourth-order Runge-Kutta integrator\,\cite{palacino-gonzalez_theoretical_2017}.
They automatically involve all possible pathways and as such are a good indicator whether the total signal should be observable in the experiment.
Moreover, the non-perturbative simulations assuming finite pulse durations give us a quantitative understanding of the nonlinear signals including transient decay times.
In short, the Schrödinger equation of the optically perturbed system is solved numerically for discrete inter-pulse delays $\tau$ yielding the final population of electronic states. Additionally, the phase of the second pulse is varied stepwise for each delay simulating the phase modulation, as described by Binz et al. for a four pulse scheme\,\cite{binz_effects_2020}.
Analog to the experiment, we sort the final-state population according to its phase signature and as such separate the different MQC channels. We use the same approach to simulate the optical 2QC by calculating $|(E_1+E_2)^2|^2$ instead of solving the Schrödinger equation, where $E_1$ and $E_2$ denote the complex-valued electric fields of our pulses.
For the simulations, we reconstruct the laser pulse in the time domain by a discrete Fourier transform of the measured pulse spectrum.
We additionally add a GDD of 420\,fs$^2$ in the spectral domain to reproduce the separately measured intensity AC. Note that higher order dispersion is neglected as it typically plays a minor role at the corresponding Fourier limit of 28\,fs.
The used transition dipole moments are 11.9\,D (Li D2), 8.4\,D (Li D1) and 3.9\,D (H$_2$Pc-He$_N$) deduced from Refs.\,\cite{heavens_radiative_1961,henriksson_semiempirical_1972}. We vary the pulse peak intensity from 0.1 to 3 GW/cm$^2$ to account for the spatially varying intensities present in our detection volume.

In Fig.\,\ref{fig4}a the time domain data for a low-intensity simulation (0.1\,GW/cm$^2$) in Li is shown.
Analog to the experimental data, we get a long-lived 1QC (not shown) and a fast decaying 2QC.
The lifetime of the simulated Li 2QC is longer than the corresponding optical 2QC, confirming the results of the experiment.
We also compare the 2QC signals to the intensity AC of the excitation pulses for GDD values of 0\,fs$^2$ and  420\,fs$^2$ (not shown).
For GDD=0\,fs$^2$, the decay time of the optical 2QC and of the intensity AC are the same while the one of the Li 2QC is longer by a factor of 1.23.
Changing to GDD=420\,fs$^2$, increases the duration of the intensity AC by a factor of 1.8.
The optical 2QC shows only minor changes (<0.5\,\%) since this signal correlates to the SHG spectrum of the pulses, which is only slightly affected by the additional GDD.
Interestingly, the Li 2QC decay time increases only by a factor of 1.09.
This means the decay time of the Li 2QC is neither strictly given by the coherence length of the excitation pulses nor by the length of their intensity overlap.
Further, the Li 2QC can either be significantly shorter (GDD=420\,fs$^2$) or longer (GDD=0\,fs$^2$) than the intensity AC.
Especially the latter case can lead to faulty assignments, as commonly pulse overlap dependent signals are expected to decay within the pulse overlap.
This discussion focuses on the half width at half maximum values of the different signals.
As an additional distinction between the signals, we observe a long lived tail in the Li 2QC (Fig. \ref{fig4}a) which is neither apparent in the optical 2QC nor in the intensity AC.

The tail in the time domain data leads to a structured line shape observable in the Fourier spectra, shown in Fig\,\ref{fig4}b.
The simulated spectral data in general show all significant
features observed in the experiment.
Note, however, that the spectral width is broader than the experimental data and the peak amplitudes and positions are slightly shifted.
A possible explanation, especially for the shift in amplitudes, are the contribution of higher intensities in the experiment.
The simulations for higher intensities (not shown) indicate a variation in the peak amplitudes and even a splitting of the peaks as soon as a saturation limit is reached.
 For a quantitative comparison of measured and simulated data detailed knowledge of the intensity distribution within the detection volume is required\,\cite{binz_effects_2020}.
 As the current experiment is conducted inside a molecular beam apparatus, the detection volume is not well confined and we omit a quantitative analysis.
 We further point out that the 2QC lineshape strongly depends on the spectrum of the excitation pulses. 
 Simplified simulations based on Gaussian pulses result in Gaussian 2QC spectra.
 Those also contain a dependency on the laser center frequency and transitions frequencies, but do not show a double peak structure, except for high intensities.

In terms of existence, duration and structured lineshape of the 2QC signal, the simulations are in agreement with the experiment. As such, we conclude that the proposed Raman-like multiphoton processes shown in Fig.\,1d can cause artificial 2QC signals in the experiments. Furthermore, evaluating the 3QC detection channel of the simulated data shows an analogous pulse overlap related signal. This indicates that the observed artificial coherences are present in any higher-order MQC signal. We did not observe any (>2)QCs experimentally as the corresponding signal amplitudes lie below our noise level.

In conclusion, we experimentally observed and simulated artificial 2QC signals in pure two level systems. The signal lifetimes and qualitative analysis indicate a pulse overlap dependent origin of the signal, however the signal lifetimes can be longer than the actual pulse overlap and their structured spectra deviate from higher-order pulse AC contributions. This may lead to misinterpretations in complex systems were genuine 2QC signals are expected. While shown here for 2QC signals, an analog behavior is expected for higher-order MQC signals. Likewise, our study is based on phase modulation/cycling to extract the MQC signals. However, the same considerations apply for non-collinear phase matching techniques and the same parasitic MQC signals are expected to contribute there as well, whenever multiple interactions with a single pulse are detected. Since the intensity scaling and the phase signature of the found artificial MQC signals are identical to the properties of genuine MQC signals, the parasitic signals should generally contribute to any MQC experiment. Typical 2DES studies would show broad spectral features that inherit the dynamics of a singly excited system. This is in agreement with the theoretical study by Rose and Krich\,\cite{rose_automatic_2021} where pulse-overlap effects were apparent for long waiting times. In principle, these features do not carry the information of the multiple excited state the experiments aim for, but of lower lying excited states. Hence, it is important to single out their contributions for an unambiguous analysis, which is especially challenging in congested spectra with in general broad features.

\begin{backmatter}
\bmsection{Funding} Deutsche Forschungsgemeinschaft (2079); European Research
Council (694965).

\bmsection{Acknowledgments} We thank Maxim F. Gelin for fruitful exchange regarding the experimental results and for providing us with the base code for the simulations.

\bmsection{Disclosures} The authors declare no conflicts of interest.

\bmsection{Data availability} Data underlying the results presented in this paper are not publicly available at this time but may be obtained from the authors upon reasonable request.

\end{backmatter}

\bibliography{Paper-AC}

\begin{thebibliography}{10}
\bibitem{note1}
In case only 3 pulses are applied, the Fourier spectrum w.r.t. t is obtained by
  detection of the signal in an optical spectrometer.
\end{thebibliography}



\end{document}